\documentclass{aastex}
\usepackage{emulateapj5}
\usepackage{apjfonts}
\usepackage{epsf}
\usepackage{timesfonts}

\def\calc{{\cal C}}
\def\calm{{\cal M}}
\def\ch{{\cal H}}
\def\dv{\Delta \upsilon}
\def\gammae{\gamma_{\rm e}}
\def\gmax{\gamma_{\rm max}}
\def\gmin{\gamma_{\rm min}}
\def\lbol{L_{\rm Bol}}
\def\ledd{L_{\rm Edd}}
\def\kms{\rm km~s^{-1}}

\def\pscl{\langle \Sigma_{\rm cl}\rangle}
\def\scl{\Sigma_{\rm cl}}
\def\pgas{P_{\rm gas}}
\def\pmag{P_{\rm mag}}
\def\rl{R_{\rm L}}
\def\rpc{R_{\rm pc}}
\def\vsh{\upsilon_{\rm sh}}
\def\nth{n_{\rm th}}
\def\nusyn{\nu_{\rm syn}}
\def\sunm{M_{\odot}}
\def\sunmyr{M_{\odot}~{\rm yr^{-1}}}
\def\tacc{t_{\rm acc}}

\def\uir{u_{\rm IR}}
\def\vorb{\upsilon_{\rm orb}}
\def\vth{\upsilon_{\rm th}}

\journalinfo{The Astrophysical Journal, 614: L21-L24, 2004, October 10}

\begin{document}

\title{Collisions among Clouds inside Dusty Torus in Active Galactic Nuclei: 
Observational Consequences}

\author{Jian-Min Wang}

\affil{Laboratory for High Energy Astrophysics, Institute of
High Energy Physics, CAS, Beijing 100039, P.~R. China.}

\slugcomment{Received 2004 July 11; accepted 2004 August 25; pulished 2004 September 15}
\shorttitle{Non-thermal Emission from Dusty Torus}
\shortauthors{WANG}

\begin{abstract}
A geometrically thick dusty torus in NGC 1068 has been unambiguously resolved by 
an infrared interferometry telescope. This implies clouds composing the dusty torus 
are undergoing supersonic collisions with each other. We show that the collisions 
form strong non-relativistic shocks, which accelerate populations of relativistic 
electrons. Torus reprocesses emission from accretion disk into infrared band. We 
show that the  energy density of the infrared photons inside the torus is 
much higher than that of the magnetic field in the clouds and the seed photons of
inverse Compton scattering are mainly from the infrared ones. The maximum 
energy of the relativistic electrons can reach a Lorentz factor of $10^5$.
We calculate the spectrum of the synchrotron and inverse Compton scattering radiation 
from the electrons in the torus. The relativistic electrons in the torus radiate 
non-thermal emission from radio to $\gamma$-ray, which isotropically diffuses in the 
region of the torus. We find the most prominent character is a peak at
$\sim 0.5-1$~GeV. We apply this model to NGC 1068 and find that the observed radio 
emission from the core component $S_1$ can be explained by the synchrotron
emission from the relativistic electrons.  We predict that there is a
$\gamma$-ray emission with a luminosity of $10^{40}$ erg/s peaking at $\sim 1$ GeV 
from the torus, which could be detected by {\em Gamma-Ray Large-Array Space 
Telescope} in the future. This will provide a new clue to understand the
physics in the torus. The non-thermal radiation from dusty torus may explain
the radio emission from Seyfert galaxies. The cosmological implications of the 
non-thermal emission to the $\gamma$-ray background radiation have been discussed.

\end{abstract}

\section{Introduction}
A geometrically thick dusty torus has been recently resolved in the well-known
Seyfert 2 galaxy NGC 1068 by an infrared interferometry telescope (Jaffe et al. 2004; 
hereafter). The dusty torus, as an essential ingredient in the unification scheme of active
galactic nuclei (AGNs), plays a key role in obscuring the broad line region (Antonucci 
\& Miller 1985; Antonucci 1993). How to maintain the thickness of the torus has long been 
puzzling in AGN physics. Krolik \& Begelman (1988) suggested that the dusty torus is composed
of discrete compressed clouds and supported by the random supersonic motion of clouds in 
$z$-direction. However, the inevitable collisions among the clouds dissipate their kinetic 
energy, an extra supply of kinetic energy to the clouds, such as a star cluster, is thus needed 
(Krolik \& Begelman 1988; Jaffe et al. 2004). What are the observational consequences of the 
collisions?

VLBA (Very Long Baseline Array) has resolved the radio components in the core region of NGC 1068.
The radio component $S_1$ at the center of the nucleus is coincident with the most
powerful infrared source in NGC 1068 (Bock et al. 2000; Jaffe et al. 2004).
The component $S_1$ should therefore relate to the dusty torus somehow physically.
The inner edge of the hypothesized torus will be ionized by UV and X-rays from 
the accretion disk and is expected to be optically thick to free-free absorption
at GHz frequencies (Neufeld et al. 1994). 
The free-free emission is suggested for the radio emission of the component $S_1$ 
from hot ionized gas of a temperature $T=6\times 10^6$K and number density  
$n=6\times 10^5$cm$^{-3}$ evaporated from dust, and the inverted spectrum 
below 5GHz in NGC 1068 is explained by the ionized gas via
free-free absorption with a total amount of
$3400\sunm$ (Gallimore et al. 2004). For typical values of the parameters, the
ablation rate is of $\dot{M}_{\rm abl}\approx 0.49\sunmyr$ via photo-ionization
(Krolik \& Begelman 1988), 
the timescale for the required mass of the ionized gas will be of $t_{\rm abl}\approx 7\times 10^3$yr. 
However, the free-free emission from hot ionized gas is proportional to the square of particle 
density and its cooling timescale  is of, $t_{\rm ff}\approx 3.2T_6^{1/2}n_6^{-1}$yr, for the 
inferred ionized gas, where $T_6=T/10^6{\rm K}$
and $n_6=n/10^6{\rm cm^{-3}}$. It is clear that $t_{\rm ff}\ll t_{\rm abl}$, namely, most of 
the ionized gas ($\sim 3400\sunm$) can not be maintained and will re-condense to dust. 
So there is no enough ionized gas to absorb the radio emission below 5~GHz in the model 
suggested by Gallimore et al. (2004). The fact that the Component $S_1$ nicely overlaps with the 
region of the most infrared bright torus (Jaffe et al. 2004) strongly implies that the
radio emission from $S_1$ is linked with the torus itself.
The explanation of this radio emission  remains open.
Does it relate to collisions among the clouds?

In this Letter, we focus on the observational consequences of the collisions among
the clouds and show that shocks due to collisions accelerate populations of relativistic 
electrons. Non-thermal emission will be radiated as observational consequences from radio 
to GeV $\gamma$-rays.
We apply this model to NGC 1068 and suggest that the radio emission of the component $S_1$ 
originates from the synchrotron emission of population of the relativistic electrons.
The predicted $\gamma$-ray emission can be detected by 
{\em Gamma-ray Large-Array Space Telescope} ({\em GLAST}).
This provides a new clue to test the mechanism supporting the torus. 

\section{Non-thermal Emission from Torus}
 
\subsection{Cloud Collisions and acceleration of electrons}
The detail micro-physics of clouds has been discussed by Krolik \& Begelman (1988). 
The clouds are supported by self-gravity and magnetic field. Random motions of
the clouds in the vertical direction support the thickness, 
$H_t\approx \dv/\Omega_{\rm K}$, where $H_t$ is the height of the torus,
$\dv$ is the random velocity in vertical direction and $\Omega_{\rm K}$
is the Keplerian velocity of the clouds at the mid-plane. 
We then have $H_t/R_t\approx \dv/\vorb$,
implying $\dv\sim \vorb$. It is expected that the clouds in torus 
are undergoing collisions in such a torus.
The clumpy torus can be described by the clumpiness $\calc=\scl/\pscl$, where
$\scl$ and $\pscl$ are the column density of each cloud and the mean column density
of the torus, respectively. Genzel et al. (1985) estimate $\calc\approx 0.1$ in
the Galactic center. The total kinetic energy dissipated by the collisions for typical
values of the parameters is given by Krolik \& Begelman (1988)
\begin{equation}
\dot{E}_{\rm diss}=1.9\times 10^{41}\rpc \Sigma_{25} 
                   \left(\frac{\vorb}{200{\rm km/s}}\right)^3~({\rm erg~s^{-1}}),
\end{equation}
where the column density of the torus $\Sigma_{25}=\pscl/10^{25}$~cm$^{-2}$
and the distance to the black hole $\rpc=R_t/1{\rm pc}$.
Part of this power is used to evaporate dust and left forms shocks which accelerate
populations of relativistic electrons. We do not tackle the energy supply to the clouds
in this paper.

The magnetic field in a cloud is expressed in term of the plasma parameter defined by
$\beta=\pgas/\pmag$,
\begin{equation}
B=1.3\times 10^{-3}\beta_{0.2}^{-1/2}\calc_{-1}^{1/2}
     \Sigma_{25}^{1/2}T_{300}^{1/2}\ch^{-1/2}\rpc^{-1/2} ({\rm Gauss}), 
\end{equation}
where $\ch=H_t/R_t$, $\beta_{0.2}=\beta/0.2$, $\calc_{-1}=\calc/0.1$ and
the temperature of the torus $T_{300}=T_t/300$K. 

The Lamor gyration radius $\rl$ of an electron with Lorentz factor $\gammae$
is $\rl=5.5\times 10^{-8}\left(\gamma_5/B_{-3}\right){\rm pc}$,
where $\gamma_5=\gammae/10^5$ and $B_{-3}=B/10^{-3}{\rm Gauss}$. This
radius is much shorter than the size of the dusty torus. The temperature of the
shocked gas during the collision is given by 
\begin{equation}
T_s=\calm^2 T_t\approx 1.2\times 10^6\dv_2^2~({\rm K}),
\end{equation}
where March number $\calm=\dv/\vth$, $\dv_2=\dv/100~\kms$ and $\vth=(kT_t/m_p)^{1/2}$ 
the sound speed. The shock velocity with respect to the unshocked gas $\vsh$ are given by
\begin{equation}
\vsh=(\Gamma+1)\left[\frac{kT_s}{(\Gamma-1)m_p}\right]^{1/2}
    \approx 2.97\times 10^7T_6^{1/2}~({\rm cm~s^{-1}}),
\end{equation}
where $\Gamma=5/3$ is the adiabatic index, $k$ the Boltzman constant,
$m_p$ the proton mass and $T_6=T_s/10^6$K the temperature of the shocked gas.
The timescale of accelerating electrons reads (Blandford \& Eichler 1987) 
\begin{equation}
\tacc=\frac{\rl c}{\vsh^2}\approx 5.8\times 10^7\gamma_5B_{-3}^{-1}T_6^{-1}~({\rm sec}).
\end{equation}
where $c$ is the light speed.

\begin{figure*}[t]
\centerline{\includegraphics[angle=-90,width=15.5cm]{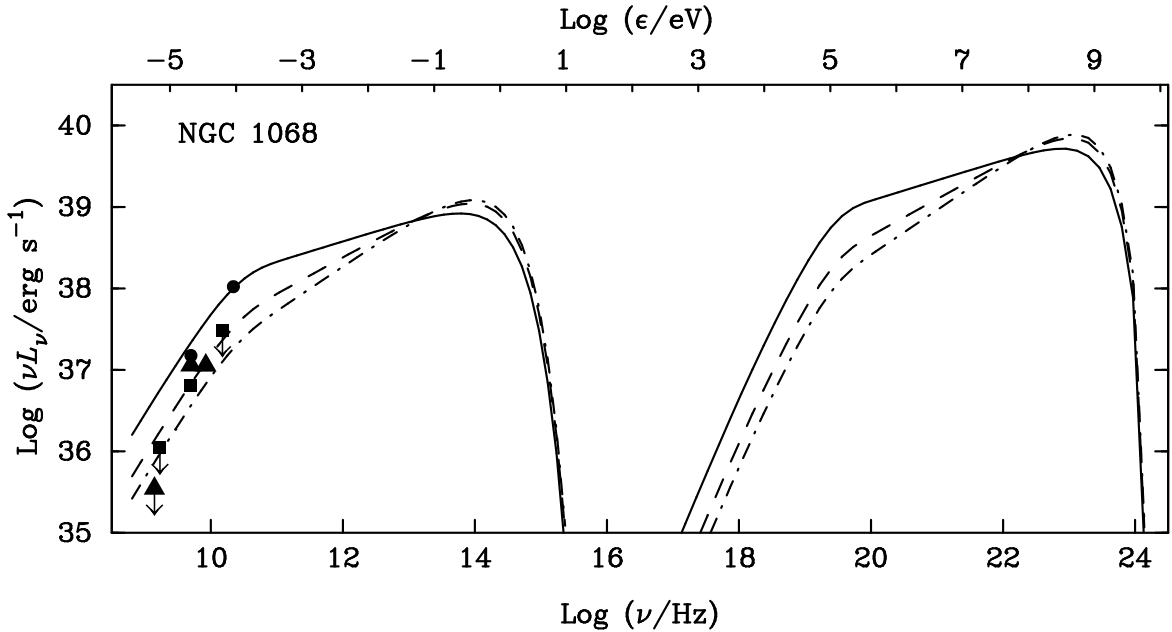}}
\figcaption{\footnotesize 
The spectrum of component $S_1$. The parameter $\beta=0.2$ is suggested in Krolik 
\& Begelman (1988). We use the typical values of $B=5\times 10^{-3}$Gauss,
$\gmax=10^5$, $\gmin=900$, $\uir=10^{-5}$erg~cm$^{-3}$ and 
$\xi_e\dot{E}_{\rm diss}=3.0\times 10^{40}$erg/s. Spectral index $\alpha$ is taken for
2.5 (solid), 2.1 (dashed) and 1.9 (dash-dotted), respectively. 
There is a clear GeV bump predicted from the present model, which reaches a level detected
by {\em GLAST}. The filled circles, squares and triangles are from Muxlow et al. (1996), 
Roy et al. (1998) and Gallimore et al. (2004). We note that the non-thermal emission from 
the torus is lower by a factor of $4\sim 5$ orders than the emission from the central engine.} 
\label{fig1}
\end{figure*}

The energy density of infrared photons inside the torus, which is reprocessed radiation from 
the accretion disk, can be estimated by
\begin{equation}
\uir=\frac{L_{\rm IR}}{2\pi^2 R_t^2\ch c}
    \approx 1.8\times 10^{-5}L_{44}\ch^{-1}\rpc^{-2}~({\rm erg~cm^{-3}}),
\end{equation}
where $L_{44}=L_{\rm IR}/10^{44}{\rm erg~s^{-1}}$ is the infrared luminosity from the 
torus. We find 
$\uir\gg u_{\rm B}$, where $u_{\rm B}$ is energy density of the magnetic field. 
Thus the cooling of the relativistic electrons is mainly due to inverse Compton
scattering of the infrared photons. The maximum energy of the
electrons will be determined by the balances between the shock acceleration and the
inverse Compton scattering of the IR photons inside the dusty torus.
The timescale of energy loss due to inverse Compton (IC) scattering is given by
\begin{equation}
t_{\rm IC}=\frac{3m_ec}{4\sigma_{\rm T}\gammae\uir}
          \approx 1.7\times 10^8 \gamma_5^{-1}L_{44}^{-1}\ch\rpc^{2}~({\rm sec}).
\end{equation}
The maximum energy reaches when the acceleration timescale is equal to the IC timescale,
\begin{equation}
\gmax\approx 1.8 \times 10^5~L_{44}^{-1/2}\ch^{1/2}\rpc B_{-3}^{1/2}T_6^{1/2}.
\end{equation}

The shock acceleration is known to generally accelerate a power law distribution of 
relativistic electrons as $\gammae^{-\alpha}$, namely, the electron distribution function
is 
%
\begin{equation}
N(\gammae)=K\gamma_e^{-\alpha}=\frac{(3-\alpha)}{\gmax^{3-\alpha}-\gmin^{3-\alpha}}
    \frac{\xi_e\dot{E}_{\rm diss}}{m_ec^2c_0\uir}\gamma_e^{-\alpha},
\end{equation}
which follows from the energy equation that the relativistic electron is getting energy
from the dissipation of the cloud collision
\begin{equation}
\int_{\gmin}^{\gmax}\dot{\gamma}_{\rm IC} m_ec^2N(\gammae)d\gammae=\xi_e\dot{E}_{\rm diss}.
\end{equation}
where $\dot{\gamma}_{\rm IC}=c_0\uir \gammae^2$ ~($c_0=3.2\times 10^{-8}{\rm erg^{-1}s^{-1}cm^{-3}}$)
is the energy loss rate of 
the electron due to inverse Compton scattering 
(Rybicki \& Lightman 1979). Here $\xi_e$ is the fraction of the thermal energy converted into the
energy of the relativistic electrons.

\subsection{Synchrotron Radiation and Inverse Compton Scattering}
For a steady torus, the cloud collisions provide a constant energy supply to relativistic 
electrons. We have shown synchrotron losses of the electrons are much less than the
inverse Compton scattering of the infrared photons. However the synchrotron radiation from
the electrons is detectable in radio band as one observational consequence.
The emergent spectrum from the torus will be given by
the synchrotron and inverse Compton scattering emission. The peak frequency of the synchrotron
emission is 
\begin{equation}
\nusyn\approx 4.2\times 10^{13}B_{-3}\gamma_5^2~({\rm Hz}),
\end{equation}
peaking at middle or near infrared band. This just overlaps with the peak frequency
of the reprocessing of the disk and thus is under-detectable.
We approximate the reprocessing of dusty torus as a black body with a single black body
temperature of $T_{\rm eff}=\left(L_{\rm IR}/2\pi^2 a R_t^2\ch c\right)^{1/4}
\approx 223 L_{44}^{1/4}\ch^{-1/4}\rpc^{-1/2}$K, where the black body radiation constant
$a=7.56\times 10^{-15}$erg~cm$^{-3}$~deg$^{-4}$. The averaged energy of the infrared photons is  
$\epsilon_{\rm IR}\approx 2.7kT_{\rm eff}\approx 5.2\times 10^{-2}L_{44}^{1/4}\ch^{-1/4}\rpc^{-1/2}$eV 
in torus, we have the peak of the inverse Compton scattering 
\begin{equation}
\epsilon_{\gamma}\approx \gmax^2 \epsilon_{\rm IR}
                 \approx 0.52~\gamma_5^2\left(\epsilon_{\rm IR}/5.2\times 10^{-2}{\rm eV}\right)
		 ~({\rm GeV}).
\end{equation}

We use the standard formulations of synchrotron and inverse Compton scattering
including Klein-Nishina effects to calculate the spectra from torus for the
typical values (Blumenthal \& Gould 1970). The ratio of energy densities of magnetic 
field to synchrotron photons is of $u_{\rm B}/u_{\rm ph}\approx 10^{2\sim 3}$ for the 
typical values of the parameters, so self-Compton scattering
emission can be neglected.  Figure 1 shows the spectra for different
electron index, but the injected energy $\xi_e \dot{E}_{\rm diss}$ is fixed.

The synchrotron spectrum beyond radio band overlaps with the reprocessed emission
from the torus itself. Since it is much lower than the later, this predicted component 
will be invisible or undistinguished. However, the component of $>100$~keV spectrum 
will be observable because the overlapped continuum from disk/corona has a cutoff at 
100keV. This is a key feature of the present model. This component
if detected only originates from the torus.
 
We emphasize that the non-thermal emission from torus is isotropic in the present model. 
The predicted spectra should be testable for any Seyfert galaxies as long as the dusty 
torus exists like in NGC 1068. In the present model, the soft X-ray emission lines
are invisible since they are too faint compared with the soft X-ray spectrum of the nucleus. 
The model of Gallimore et al. (2004) predicts strong 
features in the soft X-ray band, but obscured by the torus itself in NGC 1068. It would be 
interesting to test the model suggested by Gallimore et al. (2004) in Seyfert 1 galaxies, which 
have dusty tori and the predicted luminous soft X-ray emission lines unobscured by the tori
will be viewed by {\em Chandra}. The present model does not depend on the types of Seyfert 
galaxies provided there are geometrically thick tori in these objects. 
So it is not difficult to distinguish the two different models in principle.

\section{Application to NGC 1068}
The amount of the ionized gas can be estimated from the balance between the free-free cooling 
and the ablation of dust in the torus. 
The ablation timescale is $t_{\rm abl}=M_{\rm gas}/\dot{M}_{\rm abl}$,
where the mass of the ionized gas is $M_{\rm gas}\approx 2\pi^2 \nth m_p \ch^2 R_t^3$ and 
$\nth$ the number density of the ionized gas. We then have 
$\nth\approx 1.1\times 10^4 T_6^{1/4}\ch^{-1}\rpc^{-3/2} ~({\rm cm^{-3}})$
from $t_{\rm abl}=t_{\rm ff}$. The optical depth of free-free absorption of the ionized gas is 
$\tau_{\rm ff}=1.4\times 10^{-3}T_6^{-3/2}\nu_{\rm GHz}^{-2}n_4^2\rpc$,
where $\nu_{\rm GHz}=\nu/10^9{\rm Hz}$ and $n_4=n_{\rm th}/10^4{\rm cm^{-3}}$.
It is thus impossible for free-free absorption to cause an invert spectrum at $\sim$ GHz.
The free-free emission from the ionized gas is of 
$5.3\times 10^{32}n_4^2T_6^{-1/2}\nu_{\rm GHz}\ch^2\rpc^3$ (erg~s$^{-1})$, much lower than
synchrotron radiation from the relativistic electrons. Here the Gaunt factor is neglected in our
estimations. The thermal electrons from the evaporation has a Thomson scattering depth of
$\tau_{\rm es}=2.0\times 10^{-2}n_4\rpc$. This value is consistent with the fraction of
the polarized emission to the total due to the thermal electrons (Antounucci \& Miller 1985).
Additionally the free-free emission gets a peak flux of $10^{38}$erg/s 
at $\sim 0.1$keV, which is overwhelmed by the radiation from the core. 
We can not detect such a faint radiation component.

The Eddington ratio is roughly $\lbol/\ledd\sim 0.44$, where the mass of the black hole 
$M_{\rm BH}=10^{7.23}\sunm$ estimated from maser observation
(Greenhill et al. 1997) and the bolometric luminosity $\lbol=10^{44.98}$erg/s by 
integrating the multiwavelength continuum (Woo \& Urry 2002). It is much higher than the 
critical accretion 
rate of advection-dominated accretion flow (ADAF). This indicates that the radio emission 
from the Component $S_1$ can not originate from an optically thin ADAF in NGC 1068.  
Roy et al. (1998) suggested that the component $S_1$ may originate from synchrotron
radiation from the relativistic electrons, but it remains open how to accelerate
electrons.

\vglue 0.5cm
\begin{center}
\footnotesize
\centerline{\sc Table 1 Future Observations}
\vglue 0.1cm
\begin{tabular}{lllc} \hline \hline
Instrument & Energy Band& Threshold & Note\\
           &                           &(photons~cm$^{-2}$~s$^{-1}$)& \\ \hline
{\em INTEGRAL}/SPI$^{1}$& 20~keV-8~MeV & $2.4\times 10^{-5}$@8MeV   & $\times$\\
{\em INTEGRAL}/IBIS$^1$ & 20~keV-10~MeV & $5.0\times 10^{-5}$@10MeV & $\times$\\
{\em GLAST}$^2$         & $>100$~MeV   & $1.6\times 10^{-9}$        & $\surd$\\
{\em Agile}$^{3}$       & 30MeV-50GeV  & $5.0\times 10^{-8}$@1~GeV  & ?\\ \hline
\end{tabular}
\parbox{3.0in}
{\baselineskip 9.pt\footnotesize
\noindent
$^1$http://astro.estec.esa.nl/SA-general/Projects/Integral\\
$^2$http://www-glast.stanford.edu/mission.html\\
$^3$http://agile.mi.iasf.cnr.it/Homepage/performances.shtml
}
\end{center}
\normalsize

We use the typical values of the parameters in the present model. The dissipation rate 
of the kinetic energy via cloud collisions is $\dot{E}\approx 2.0\times 10^{41}$ergs/s, 
and $\xi_e=0.15$. The 
radiation luminosity via inverse Compton scattering is of $\sim 10^{40}$ergs/s.
We find that the radio spectrum of $S_1$ can be generally fitted by synchrotron
emission. 
The most prominent character in Figure 1
is the GeV bump, which is caused by inverse Compton scattering of the infrared photons
inside the torus. This bump corresponds to a flux of 
$\sim 10^{-9}$ photons~cm$^{-2}$~s$^{-1}$ at 1~GeV and 
$\sim 10^{-8}$ photons~cm$^{-2}$~s$^{-1}$ at $>100$MeV.
Detection of this component is essential to probe physics inside the torus. 

Seyfert galaxies have been observed by {\em EGRET}, showing upper limit with 2$\sigma$
(Lin et al. 1993). Table 1 lists the sensitivities of several missions. The last column
gives the comments to observations. We find that the Italian mission {\em Agile} could marginally
detect GeV bump in NGC 1068 and {\em GLAST} has capability of detecting this GeV bump. 

The synchrotron self-absorption (SSA) depth is numerically given by
$\tau_{\rm syn}\approx 1.5\times 10^{-5}K_{56}\rpc^{-2}$ ($\alpha=2.5$, $\gamma_{\rm min}=900$,
$B=5.0\times 10^{-3}$Gauss) at $1$ GHz, where $K_{56}=K/10^{56}$, for a spatially homogeneous 
distribution of the relativistic electrons inside the torus. This shows less importance of the SSA,
however we note that this may severely underestimate $\tau_{\rm syn}$ depending on the spatial 
density distribution of the electrons. This is because 
we use the  number density of the relativistic electrons averaged
over the entire torus, which is much lower than the local. Additionally, the GHz spectrum 
may rely on the break energy of the electron distribution, which is determined by the detail of 
processes of the acceleration and cooling. We remain it as a free parameter to fit the spectrum 
since we pay main attention to the observational consequences of cloud collisions. Future detail 
numerical simulations of cloud collisions may display some details of accelerations so that the
minimum energy of the electrons and SSA depth will be given.

\section{Discussions}
Collisions simultaneously lead to loss of angular momentum of molecular clouds and then determine the
accretion rate of the black hole (Krolik \& Begelman 1988). Therefore the present model predicts
a strong correlation between the radio emission from torus and a parameter representing the accretion 
rate. Indeed Ulvestad \& Ho (2001) find a very strong correlation, 
$P_{\rm 6cm}\propto L_{\rm [O~III]}^{\sim 1.5}$ (estimated from their Fig 5) in the Palomar Seyfert 
galaxies, where $L_{\rm [O~III]}$ is the [O {\sc iii}] luminosity and $P_{\rm 6 cm}$ is the radio 
emission power at 6cm. $L_ {\rm [O~III]}$ is a good indicator of the accretion rate of the black hole 
since $L_ {\rm [O~III]}\propto L_{\rm ion}$, where $L_{\rm ion}$ is the ionizing luminosity.
We note that the collisions lead to angular momentum loss of clouds, which will be
accreted onto the black hole. Therefore the radio emission and $L_{\rm [O III]}$ are
related with the physics of cloud collisions. The $P_{\rm 6cm}-L_{\rm [O III]}$ correlation could be 
explained by the present model. The predicted radio flux in the present  model depends on electron index 
$\alpha$, a more careful study on this will be carried out in future.

The isotropic radiation of the relativistic electrons may imply a significant contribution to the 
$\gamma$-ray background radiation. If the typical $\gamma$-ray luminosity from torus is of
$L_{\gamma}^{\rm AGN}\approx 10^{40}$erg/s, we have the contributed flux to the $\gamma$-ray
background radiation $F_{\gamma}\approx \phi_{\rm AGN}L_{\gamma}^{\rm AGN} \Delta V/4\pi d_L^2$,
where $\phi_{\rm AGN}$ is AGN number density peak, $\Delta V$ is the shell volume of the
AGN number peak at redshift $z\approx 1$ and $d_L$ is the distance the peak shell. 
We have the contribution of AGN torus to the $\gamma$-ray background 
$F_{\gamma}\approx 0.6~{\rm keV~cm^{-2}~s^{-1}~sr^{-1}}$ for 
$\phi_{\rm AGN}=5.0\times 10^{-5}~{\rm Mpc^{-3}}$ from X-ray observation 
(Steffen et al. 2003) and the peak shell width $\Delta R\approx d_{\rm L}$.
The deduced $\gamma$-ray background emission is of $1.0~{\rm keV~cm^{-2}~s^{-1}~sr^{-1}}$
below 1~GeV (see Figure 1 in Loeb \& Waxman 2000). The predicted GeV flux 
significantly contributes to background emission below 1~GeV.
It is interesting to note that the background emission spectrum
shows increases below 0.6~GeV (Loeb \& Waxman 2000). It is worth investigating
this subject in detail further. 

\section{conclusions}
We suggest that the collisions among the clouds inside the torus produce populations of 
relativistic electrons in active galactic nuclei. Non-thermal emission from relativistic 
electrons is detectable in radio and $\gamma$-ray band.  The radio emission from the Component
$S_1$ in NGC 1068 can be explained by the present model. It is predicted that
there is a GeV bump in NGC 1068, which is caused by inverse Compton scattering off the infrared
photons inside dusty torus. This GeV component can be detected by {\em GLAST}. We also show that 
the GeV bump of radiation from torus could significantly contribute to the background radiation
of $<1~{\rm GeV}$. 
 
\acknowledgements
I thank the anonymous referee for the helpful and prompt comments. 
This research is supported by Grant for Distinguished Young Scientists from
NSFC, Hundred Talent Program of CAS and 973 project.

\end{document}